\documentstyle[aps,twocolumn,prb,epsf]{revtex}

\begin{document}

\def\be{\begin{equation}}
\def\ee{\end{equation}}
\def\beq{\begin{eqnarray}}
\def\eeq{\end{eqnarray}}
\def\lsim{\:\raisebox{-0.5ex}{$\stackrel{\textstyle<}{\sim}$}\:}
\def\gsim{\:\raisebox{-0.5ex}{$\stackrel{\textstyle>}{\sim}$}\:}
\def\d{\displaystyle}
\def\u{\underbar}

\draft

\twocolumn[\hsize\textwidth\columnwidth\hsize\csname @twocolumnfalse\endcsname

\title{On the Normal state electronic properties of Layered $Sr_{2}RuO_{4}$.
 }

\author{M S. Laad and E. M\"uller-Hartmann }
\address{$^1$Institut f\"ur Theoretische Physik, Universit\"at zu K\"oln,
Z\"ulpicher Str 77, 50937 K\"oln, Germany}

\maketitle

\begin{abstract}
  Based on a comprehensive perusal of experimental results, we construct a 
microscopic model describing the normal state of
layered $4d$ oxide
$Sr_{2}RuO_{4}$ incorporating relevant quantum chemical features of
the material.  The high-$T$ anomalies are explained within a 
Luttinger liquid (LL) picture.  Interlayer one-particle hopping drives a dimensional
crossover to a correlated Fermi liquid below a scale $E^{*}<<t_{\perp}$.  Using recently developed chain-dynamical mean field theory, the low
value of
 $E^{*}$, as well as various puzzling features of the low-$T$ normal state are
explained as manifestations of the crossover from the
 high-$T$ Luttinger liquid state 
in a consistent way.       
\end{abstract}
     

]

 Ruthenates constitute a class of $4d$ transition metal (TM) oxides existing
in cubic as well as layered forms, and show a range of properties, 
from a ferromagnetic bad metal with a non-Fermi-liquid response (in
$SrRuO_{3}$), via unconventional superconductivity (SC)
 in layered $Sr_{2}RuO_{4}$,
to metamagnetic (quantum criticality?) in $Sr_{3}Ru_{2}O_{7}$~\cite{[1]} .   
These findings have led to intense activity to understand correlation effects
in $4d$ TM oxides, in particular, the 
unconventional SC of $Sr_{2}RuO_{4}$~\cite{[2]}.

  $Sr_{2}RuO_{4}$ with $Ru$ in a $4d^{4}$ state and a SC transition
at $T_{c}\simeq 1.5K$~\cite{[3]}, is  
isostructural to pure $La_{2}CuO_{4}$, however, {\it without} buckling
of $RuO_{2}$ planes.  The crystal field breaks the first Hund's rule, resulting
in spin $S=1$ at each $Ru$ site.  The tetragonal BCT structure seems to remain undistorted 
to the lowest $T$ (the $RuO_{6}$ octahedra are elongated, leading to a lifting
of the three-fold $t_{2g}$ degeneracy).  
 On structural and chemical grounds,
one infers three electronically active
 bands: one ($d_{xy}$) two-dimensional band and two 
($d_{yz,zx}$) 1D bands.  These are indeed seen in band structure 
calculations~\cite{[4]}, which give a Fermi surface seemingly in agreement with 
dHvA results~\cite{[5]}.  Further, based upon early ideas   
~\cite{[6]}, these calculations attempted to search for proximity to
a ferromagnetic instability.  Instead, proximity to 
{\it incommensurate} magnetic order (due to the almost nesting character of the
1D ($d_{yz,zx}$) bands was found, in nice agreement with recent inelastic neutron scattering (INS) results~\cite{[7]}.

  The agreement between LDA calculations 
and experiment stops as soon as finite energy/temperature responses are probed.
At low-$T$ (i.e $T_{c}<T<T^{*}=30K$), the system shows all the characteristics of a 
correlated Fermi liquid (FL)~\cite{[8]}.  The $\gamma$-coefficient of the low-$T$
electronic specific heat yields a fourfold enhancement of the effective mass 
over the LDA prediction.  Integrated 
photoemission spectra at low-$T$ are quite different from the LDA results, but
are well reproduced by inclusion of a dynamical correction (to second order)
to the one-particle self-energy.  The dc resistivity~\cite{[8]} shows anisotropic FL 
form, with a low residual value just above $T_{c}$.  Careful perusal 
shows, however, that the ratio of "Woods-Saxon ratios" (along $c$ and $ab$ directions), $(A_{c}/\gamma^{2}):(A_{ab}/\gamma^{2}) \simeq 1000$, a value too large to be related to a ratio of Fermi surface areas, indicating subtle influence of correlation effects also at lower $T$.
  Optical measurements~\cite{[9]} 
give more evidence for anisotropic, correlated FL behavior at low-$T$.

  Very interesting and unusual changes occur as $T$ is raised above a 
characteristic scale $T^{*}\simeq 30$K.  A smooth crossover to a non-FL metallic state
(which shares many similarities with the non-FL normal
state of underdoped HTC cuprates) is clearly revealed by experiment.  
The anisotropic dc resistivity shows the first sign of this change: $\rho_{ab}(T) \simeq A+BT$ for $T^{*}<T<900$K, while $\rho_{c}(T) \simeq C_{1}+C_{2}T$ for
$T^{*}<T<120$K, with a crossover to an insulator-like form for $T>120$K~\cite{[9]}.
Clearly, the system falls into the "bad metal" class inspite of its low
resistivity near $T_{c}$.  
Interesting facts are revealed by optical studies~\cite{[9]}, where an extended 
"Drude" fit was used to analyze the low-energy response.  Both
the $c$-axis scattering rate and effective mass were found to 
have strong $\omega,T$ dependences.  While $\tau_{c}^{-1}(\omega) \simeq a+b\omega$, the effective mass
  $m_{c}^{*}(\omega)$ has a near-logarithmic dependence for $T>T^{*}$.
Below 30K, $m_{c}^{*}$ increases to 
40 times its band value at small energies, while $\tau_{c}^{-1}(\omega) \simeq
a'+b'\omega^{2}$.  However, the corresponding inplane $\tau_{ab}^{-1}(\omega)$ and $m_{ab}^{*}(\omega)$ are weakly $\omega$-dependent, pointing
to a direct correlation of the change of the electronic state with the change
in $c$-axis dynamics.

  Most interesting is the information obtained from recent INS 
studies~\cite{[7]}, where a detailed analysis of
the spin dynamics was carried out. 
 At low-$T$, the magnetic fluctuations are found to be
dominated by incommensurate peaks related to the Fermi surface nesting of the
two quasi-1D bands.  A pronounced softening in the spectrum near
${\bf Q_{i}}=(0.34, 0.34, 0)$ occurs upon cooling.  Additionally, near ${\bf Q_{i}}$ and for $T>T^{*}=30$K, $\chi^{"}({\bf q},\omega)$ shows the so-called
$\omega/T$-scaling: $\chi^{"}({\bf q},\omega) \simeq \omega^{-\gamma}f(\omega/T)$ with $\gamma=0.75$ 
and $f(x) \propto x$ as $x\rightarrow 0$, reminiscent of similar behavior observed in certain rare-earth
based heavy fermion compounds near their quantum critical points, but there
 essentially over the full Brillouin
zone~\cite{[10]}. 
Hence the anomalous $\omega/T$ scaling found only near ${\bf q}={\bf Q_{i}}$
in $Sr_{2}RuO_{4}$ must be ascribed to the proximity to the incommensurate AF instability involving 
the almost nested 1D bands.  At low-$T$, however, a conventional Lorentzian form is 
adequate.  Finally, the NMR relaxation rate shows a noticeable $T$ dependence
above $60$K, with Korringa like behavior at lower $T$ 
along with a sizable anisotropy ($\simeq 3$), implying an easy-axis spin
 anisotropy~\cite{[7]}.  Thus these
 results, taken together with~\cite{[10]} , reveal a {\it smooth} crossover from an
anomalous high-$T$ bad-metallic to a low-$T$ anisotropic, renormalized FL state 
across $T^{*}$.

  In this letter, we provide a unified understanding of these high-$T$ anomalous features along with a 
description of the crossover to a FL using a material specific starting model.  In particular, we show
how specific quantum chemical features along with ubiquitous strong correlations are necessary ingredients
for a unified understanding of the normal state.  This is important because a detailed understanding of
the nature of the normal state is crucial to pinpoint the "glue" that drives superconductive pairing.

  Werner~\cite{[11]} has recently discussed the normal state physics of $Sr_{2}RuO_{4}$ from a similar
starting point.  Nevertheless, in what follows, we provide a very different theoretical modelling of the 
normal state that ties together all the above experimental results within a single approach.

  To start, we notice that in $Sr_{2}RuO_{4}$, electron hopping involves Ru-$4d$-O-$2p$ hybridization, which has a peculiar structure because of geometrical constraints imposed by lattice 
structure and $t_{2g}$ orbital orientation.  Considering an isolated $RuO_{2}$ plane to start with, it is
easy to show~\cite{[12]} that the $4d_{xy}-2p_{x,y}$ hybridization leads to a 2D band, and the 
$4d_{yz,zx}-2p_{x,y}$ hybridization leads to two 1D bands, as found in LDA studies.  In the undistorted,
BCT structure, and without direct $O-O$ hopping, these three form mutually non-hybridizing bands.  Small
corrections arise from direct $O-O$ hopping: we do not consider them to begin with.  This leads to an 
important conclusion: there is {\it no} interband mixing at 
 one-particle level.

  Local coulomb interactions, like the interorbital coulomb interaction $U_{\mu,\mu'}^{\sigma\sigma'}$ as well as the 
Hund's rule coupling $J_{H}$ do lead to interband scattering processes.  Finally, 
in $Sr_{2}RuO_{4}$, the spin-orbit ($s-o$) coupling is important, as shown 
by the magnetic anisotropy~\cite{[7]}.  This lifts the
 d-$yz,zx$ orbital degeneracy and leads to a small mixing of the $t_{2g}$ orbitals. 
  So in the situation where the $s-o$ coupling (small interband one-particle
hybridization) is irrelevant, 
the 1D correlations will dominate the physical response.  

  Strictly speaking,
one should also consider the 2D-$d_{xy}$ band in the analysis, but, since the 
singular effects of interactions are much stronger in 1D than in 2D (especially
at fillings far away from commensurability), we do not consider the $d_{xy}$
band at this point.  However, it will become 
important once the dimensional 
crossover (see below) occurs at lower $T$.

  Additional perturbations like interlayer one-particle hopping will  lead to a dimensional crossover, restoring correlated FL behavior.  We will consider 
these effects later below.  The Hamiltonian describing an isolated $RuO_{2}$ 
plane is thus written as~\cite{[11]},

\be
H=\sum_{<ij>,\mu,\mu',\sigma}t_{\mu\mu'}(c_{i\mu\sigma}^{\dag}c_{j\mu'\sigma}+h.c) + H_{int}
\ee
where

\be
H_{int}=\sum_{i\mu\mu'\sigma\sigma'}U_{\mu\mu'}^{\sigma\sigma'}n_{i\mu\sigma}n_{i\mu'\sigma'} - \sum_{i,\mu,\mu'}{\bf S_{i\mu}}{\cdot}(J_{H}{\bf S_{i\mu'}} -\lambda{\bf L_{i\mu}}).
\ee
where $\mu, \mu'$ represent the $t_{2g}$ orbitals $xy,yz,zx$, with $U_{\mu\mu'}^{\sigma\sigma'}=U\delta_{\mu\mu'}+(1-\delta_{\mu\mu'})[U_{1}\delta_{\sigma\sigma'}+U_{2}\delta_{\sigma,-\sigma'}]$
 and the corresponding hopping
parameters are taken from tight-binding fits to LDA results~\cite{[4]} , but $U, U_{\mu,\mu'}^{\sigma\sigma'}, J_{H}$ and $\lambda$ must be regarded as parameters obtained by fitting to high-energy spectroscopic and magnetic data.

  The bandstructure described above is modified in the presence of spin-orbit 
coupling.  Indeed, the $s-o$ term couples states with different $S_{z}$.  
Making a unitary transformation to new operators which create (destroy) holes 
in the renormalized ground state containing contributions from the excited
states of the original model via $s-o$ matrix elements, 
the hopping becomes spin-dependent, parametrized by:
$t_{\mu\mu'}^{\sigma\sigma}=t, t_{\mu\mu'}^{-\sigma\sigma}=t\eta$, where $\eta$
is treated as a parameter to fit the available magnetic data.  
The hopping part now reads,

\be
H_{0}=\sum_{<ij>,\mu,\mu',\sigma\sigma'}t_{\mu\mu'}^{\sigma\sigma'}(C_{i\mu\sigma}^{\dag}C_{j\mu'\sigma'}+h.c)
\ee
where $C_{i\sigma}=\sum_{\sigma'}V_{\sigma\sigma'}c_{i\sigma'}$, with $V_{\sigma\sigma'}$ a 2x2 unitary matrix.
This rotation in spin space also produces new (four fermion) interaction 
terms in $H_{int}$ above.  However, these do not 
modify the special features of the hybridization, and so leave the low
energy physics qualitatively unaffected.  They should, however, be included in
a strictly rigorous formulation.      

  From the above discussion, the dominant physical effects  
arise from the two 1D ($d_{zx,yx}$) bands, which are coupled to the 2D 
$d_{xy}$ band by $U_{\mu,\mu'}^{\sigma\sigma'}, J_{H}$.  From the structure of
the hopping matrix, the d-$xy,yz,xz$ electrons do not hybridize 
with each other but interact via $U_{\mu,\mu'}^{\sigma\sigma'}, J_{H}$, giving
rise to additional strong scattering processes at low energy.  This is an 
additional factor playing an important role in our description of the 
normal state.
 
  Finally, to describe the 1D-3D
crossover (which we will do later), we notice that the dominant coupling between the $RuO_{2}$
layers occurs via a one-particle tunnelling matrix element, $t_{\perp}$ 
between the 1D bands, since the $d_{xy}$ orbitals have negligible overlap
between neighboring layers.
  We mimic~\cite{[11]} the effect of finite $J_{H}$ by setting $U_{\mu\mu'}^{\sigma,-\sigma}=U_{1}>U_{2}=U_{\mu\mu'}^{\sigma\sigma}$.  

  To bosonize the model represented by Eqs.(2-3),
  we notice that each of the (non-degenerate in presence of $s-o$
coupling) 1D ($d_{yz}$ for e.g.) bands with 
(equi-orbital) 1D hopping and $U$ is modelled by a LL
Hamiltonian with average band filling $n=2/3$.  This means that the 1D
charge correlations can be modelled by a gaussian model for the charge bosonic
field $\phi_{\rho}(x)$, since umklapp scattering is inoperative away from half-filling, and backscattering renormalizes to zero~\cite{[13]} for repulsive interactions.
  In addition, the
inter-orbital interaction term $U_{1}n_{id}^{yz}n_{id}^{zx}$ acts like a 
strong scattering potential for the $yz$ band carriers in the limit where
the interband hybridization $t_{yz,zx}$ is effectively zero (a 
similar term comes from $U_{1}n_{id}^{xy}n_{id}^{zx}$). There is an analogous 
effect coming from $U_{2}$.  So 
in addition to the 1D LL physics 
in each of the $yz,zx$ channels, one 
has to treat the infra-red singular effects (in 1D)
 arising from $U_{1}$ and $U_{2}$~\cite{[14]}.  

  In the 1D case, spin-charge separation allows us to write down separate 
bosonized Hamiltonians for spin and charge collective modes.  In our case 
relevant to $Sr_{2}RuO_{4}$, the final bosonized model for the $zx$ band reads,

\be
H_{c}^{(0)}= \int dx [K_{\rho}\Pi_{\rho}^{2}+K_{\rho}^{-1}(\partial_{x}\phi_{\rho}(x))^{2}]
\ee
with $H_{xre}=(U_{1}/(\pi a)^{2})\int dx n_{d}^{yz}\partial_{x}\phi_{\rho}(x)$ from the inter-orbital interaction as described above.
For the $yz$ band, the same holds true with $x\rightarrow y$.  These equations
describing the 1D charge correlations are precisely those of "shifted"
collective charge-density oscillations, and leave the LL behavior
unchanged.  They do, however, lead to an important rescaling of the Fermi-edge
singularity (FES) exponent observed in knockout experiments with hard X-rays~\cite{[14]}. 
  
  From the spin-dependent hopping above, one infers that the 
magnetic correlations are described by an effective anisotropic spin chain:

\be
H_{s}=\sum_{<ij>}[J_{z}S_{i}^{z}S_{j}^{z} + (J_{xy}S_{i}^{+}S_{j}^{-} + J_{\perp}S_{i}^{+}S_{j}^{+}+h.c)]
\ee
with $J_{z}=4t^{2}(1+\eta)^{2}/U, J_{xy}=4t^{2}(1-\eta^{2})/U$ and
$J_{\perp}=4t^{2}\eta^{2}/U$ where the  
anisotropy parameter, $\Delta=(J_{xy}/J_{z})=(1-\eta)/(1+\eta)$.  The bosonized 
version of this model has the two-cosine structure of the XYZ model:

\be
H_{s}=H_{s}^{(0)} 
-\int[\frac{m}{a\pi}cos(\beta_{s}\Phi_{s}) - 
\frac{m'}{a\pi}cos(\beta_{s}'\Theta_{s})]dx.
\ee
where

\be
H_{s}^{(0)}=\frac{u_{s}}{2}\int dx[(\Pi_{s}^{2}(x)+(\partial_{x}\phi_{s})^{2})
\ee
  This has a duality property~\cite{[13]}: when $J_{z}>|J_{xy}|-|J_{\perp}|$, $J_{xy}$
scales down to zero, while $J_{z}, J_{\perp} \rightarrow\infty$.  The dual field
 $\Theta_{s}$ gets ordered, leading to dominant $SDW^{xy}$ correlations.  
When $J_{z}<|J_{xy}|-|J_{\perp}|$ (which for our case means $-2/3<\eta<0$)
$J_{\perp}$ scales to irrelevance, and the resulting picture is qualitatively 
the same as for the XXZ model, and $SDW^{z}$ correlations are dominant.
In our case, we estimate $K_{\sigma}=0.87$.         
In this regime,       
the SDW correlation function will be determined by{\it both} $K_{\sigma}$
and the 
gapless {\it charge} sector, i.e., by $K_{\rho}$.

  The spin correlation function is 
evaluated as a statistical average over phase variables~\cite{[15]} , giving 
$\chi_{s}^{zz}(x)=<{\bf S}(x){\cdot}{\bf S}(0)> \simeq \frac{cos(Q_{i}x)}{x^{K_{\sigma}^{-1}+K_{\rho}}}$
for the equal-time part.  At finite $T$, and near ${\bf q}={\bf Q}_{i}$, one obtains 
$\chi^{zz}(Q_{i},T) \simeq T^{-K_{\sigma}^{-1}+K_{\rho}}$.
The transverse dynamical spin susceptibility is given by

\be
\chi_{\perp}({\bf q},\omega) \simeq \frac{A_{\perp}}{T^{K_{\sigma}^{-1}-K_{\rho}}}\frac{\omega}{T}
\ee
and it goes like $\chi_{\perp}({\bf q},\omega) \simeq T^{-\gamma}\frac{\omega}{T}$
for $q'=(q-Q_{i}) \simeq 0$, with $\gamma=(K_{\sigma}^{-1}-K_{\rho})$.   
The NMR relaxation rate, $1/T_{1}$ follows directly as

\be
\frac{1}{T_{1}}=\frac{T}{\omega}\sum_{\bf q}\chi"({\bf q},\omega) 
\propto T^{-\gamma}.
\ee

Finally, to compute the value of the charge stiffness $K_{\rho}$,
we consider the Hamiltonian

\be
H_{c}=H_{c}^{(0)}+\frac{U_{1}}{(a\pi)^{2}}\int (n_{d}^{xy}+n_{d}^{yz})\partial_{x}\phi_{\rho}(x) dx.
\ee
where $u_{\rho}K_{\rho}=v_{F}$ and $u_{\rho}/K_{\rho}=v_{F}+U$.  In the
intermediate coupling regime, $U \simeq 2.1$eV, $v_{F}=0.7$eV~\cite{[4]}, and $K_{\sigma}=0.8$, one gets
$K_{\rho} \simeq 1/2$
giving $\gamma=0.65$ and the LL Fermi surface exponent,
$\alpha=1/8$.  On the 
other hand, $K_{\rho}=0.4$ (corresponding to $U\simeq 2.4$eV) 
yields $\gamma=0.75$ and $\alpha=0.23$.
The value $\gamma=0.75$ gives good fits to the quantum-critical 
scaling behavior of $\chi"({\bf q=Q}_{i},\omega)$ above $T^{*}$.   

  The NMR relaxation rate is then $1/T_{1} \propto T^{-0.75}$ which is not 
inconsistent with the observed $T$-dependence for $T \geq 100$K.  Eqn.(7) with
$\gamma=0.75$ is also completely consistent with the experimental result, 
however, for $T>40-50$K.  At lower-$T$, $\chi(0,T)$ levels off and approaches 
a constant, consistent with a crossover to correlated FL behavior~\cite{[16]}.  
The in-plane dc resistivity shows almost linear behavior
for $50K < T < 900$K, consistent with the Luttinger liquid physics above 50 K.
More support for this picture comes from optics~\cite{[17]}  where a crossover from incoherent response to a Drude-like response is indeed observed as $T$ is 
lowered; however, interestingly, the scattering rate extracted from a generalized Drude fit shows $T^{2}$
dependence only up to $\simeq 30$K, in agreement with INS results.
               
  To consider the LL-FL crossover as a function of $T$, we 
notice that the interlayer hopping, $t_{\perp} \simeq 0.02$eV (200K)~\cite{[4]}.
A description of the effect of $t_{\perp}$ requires consideration of a model 
with coupled $RuO_{2}$ layers.  Since the interlayer hopping for the
$d_{xy}$ band is much smaller than for the $d_{yz,zx}$
bands in the undistorted BCT structure, we are led to consider the
 model of coupled chains:

\be
H=\sum_{\nu}H_{1D}^{\nu} - \sum_{i,\nu,\nu',\sigma}t_{\perp}(C_{i\nu\sigma}^{\dag}C_{i\nu'\sigma}+h.c)
\ee
for each of the 1D ($\nu=yz,zx$) bands.  A description of the crossover by 
perturbation theory in $t_{\perp}$ is beset with difficulties, and is valid in
the LL regime, but fails to reproduce the FL regime.  Perturbation approaches in
 interaction work in the FL regime, but fail in the LL regime.  An attractive
way out is provided by a recent non-trivial extension of dynamical mean-field
theory (DMFT) that replaces a single site by a single chain connected via 
$t_{\perp}$ to $z_{\perp}$ nearest neighbors, with $z_{\perp}\rightarrow\infty$~\cite{[18]}.
  Rigorously, one needs a numerical solution  
for the {\it full} single-chain propagator $G({\bf k},\omega)$, which is a very hard
task, when considered together with the usual DMFT selfconsistency~\cite{[18]}.
Fortunately, several conclusions can still be drawn without attempting a full 
solution.  In the 1D regime, the in-chain self-energy, $\Sigma(k,\omega)\simeq t((k,\omega)/t)^{1/(1-\alpha)}$.  From the DMFT equations, $t_{\perp}$
becomes relevant when $t_{\perp}>\Sigma$, yielding the crossover scale,
$E^{*} \simeq t_{\perp}(t_{\perp}/t)^{\alpha/(1-\alpha)}$.  In our case, this 
gives $E^{*} \simeq 60$K, qualitatively in agreement with, but somewhat higher
 than $T^{*} \simeq 30-40$K from 
experiment.  
  
  At $T<E^{*}$, chain-DMFT leads to anisotropic FL behavior.
In particular, when $t_{\perp}<<t$ and at low energies, all one-particle quantities
 obey the scaling $\omega'=\omega/E^{*}, k'=kE^{*}/t$ and $T'=T/E^{*}$; i.e.,
$t\Sigma(k,\omega,T)=E^{*}t_{\perp}\Sigma'(k',\omega',T')$ and $tG(k,\omega,T)=
(E^{*}/t_{\perp})G'(k',\omega',T')$ where $\Sigma$ and $G$ are universal 
functions associated with the crossover.  A low-frequency expansion of $\Sigma$
in the FL regime gives the quasiparticle residue $Z \simeq (t_{\perp}/t)^{\alpha/(1-\alpha)}=E^{*}/t_{\perp}$.  Unlike in a conventional FL, this bears no 
resemblance to the effective mass enhancement, since both $(\partial\Sigma/\partial k)$ and $(\partial\Sigma/\partial\omega)$ scale in the same way.  
The interchain resistivity, $\rho_{\perp}(T)/\rho_{0}=(t/E^{*})R(T/E^{*})$ with
 $R(x<<1) \propto x^{2}$ and $R(x>>1) \propto x^{1-2\alpha}$.  
And the resistivity enhancement,
$\rho_{\perp}(T)/ \rho_{0} =A (T/t)^{2}$ with $A=(t/t_{\perp})^{3/(1-\alpha)}$.
The resulting anisotropy of the Woods-Saxon ratio, 
$A_{c}/A_{ab}=(a/c)^{2}A \simeq 1000$ 
for $\alpha=0.23$, which is indeed in the right range~\cite{[9]}.
Finally, the $c$-axis optical response is incoherent above $E^{*}$,
with a coherent feature carrying a relative weight $\simeq Z^{2}$ appearing at 
low-$T$, again in qualitative agreement with observations~\cite{[10]} . 
An obvious inference from the above is that increasing $T$ should lead to a 
disappearance of the quasicoherent features in photoemission.  This may already have been observed experimentally~\cite{[19]}. 

  At low-$T$, in the correlated FL regime, usual DMFT should provide a 
consistent description of electronic correlations.  Such a program has been implemented~\cite{[20]}  for the multiband system of the $t_{2g}$ bands coupled by 
$U_{\mu\mu'}^{\sigma\sigma'}, J_{H}$.  With $U=2.5$eV, $J_{H}=0.4$eV, the 
effective mass enhancement (from the self-energy) is $m^{*}/m \simeq 3-4$, 
completely consistent with specific heat data and dHvA measurements.  In our 
picture, therefore, the effective mass enhancement arises from renormalization
effects caused by local electronic correlations (DMFT) and has a very different
 origin from the one proposed by Werner~\cite{[11]}.  
Interestingly,
this approach~\cite{[20]} also reconciles the apparent conflict between 
ARPES and dHvA data at low-$T$.

   To conclude, starting from a material specific model for the layered 
TM oxide $Sr_{2}RuO_{4}$, we have described how the various high
$T$ anomalous features can be understood within Luttinger liquid ideas.  
The effects of $s-o$ coupling, necessary to obtain consistency with magnetic data, are consistently incorporated, leading  
to a new modelling for the high-$T$ LL state compared to that
 of~\cite{[11]}.  Finally, 
using the Luttinger liquid exponents obtained there, a {\it smooth} crossover to an anisotropic, 3D, correlated FL metallic state (below $E^{*}=60$K) 
is obtained within the recently developed chain-DMFT.  The low-$T$
 specific heat and
static spin susceptibility are enhanced by conventional FL renormalization below
 $T^{*}$.  Various thermodynamic and transport properties, some inexplicable 
within conventional scenarios, find a consistent explanation 
as manifestations of the interplay between the high-$T$
LL state (irrelevance of $t_{\perp}$) and the low-$T$ correlated FL state.


  We thank M. Braden for fruitful discussions and R. Werner for preprints.
This work is performed within the research program of the Sonderforschungsbereich 608 of the Deutsche Forschungsgemeinschaft.

\end{document}